\documentclass[journal]{IEEEtran}
\usepackage{amsmath,amsfonts,amssymb}
\usepackage{amsthm}
\usepackage{graphicx}
\usepackage{float}
\usepackage{cite}
\usepackage{url}
\usepackage{booktabs}
\usepackage{xcolor}
\usepackage{hyperref}
\usepackage{mathtools}
\usepackage{array}
\usepackage{multirow}
\usepackage{tabularx}
\newtheorem{theorem}{Theorem}

\newtheorem{definition}{Definition}
\newtheorem{proposition}{Proposition}

\DeclareMathOperator{\argmax}{arg\,max}
\DeclareMathOperator{\argmin}{arg\,min}

\title{Game-Theoretic Resilience Framework for Cyber-Physical Microgrids using Multi-Agent Reinforcement Learning}

\author{S Krishna Niketh,
        Sagar Babu Mitikiri,
         V Vignesh, Vedantham Lakshmi Srinivas, Mayukha Pal*
        {}
\thanks{*(Corresponding author: Mayukha Pal, e-mail: mayukha.pal@in.abb.com)}
\thanks{Mr. S. Krishna Niketh is a Data Science Research Intern at ABB Ability Innovation Center, Hyderabad 500084, India, and also an undergraduate student at the Department of Electrical Engineering, Indian Institute of Technology Tirupati, 517619, India.}
\thanks{Mr. Sagar Babu Mitikiri is a Data Science Research Intern at ABB Ability Innovation Center, Hyderabad 500084, India, and also a Research Scholar at the Department of Electrical Engineering, Indian Institute of Technology
(ISM), Dhanbad 826004, India}
\thanks{Dr. V. Vignesh is an Asst. Professor with the Department of Electrical Engineering, Indian Institute of Technology, Tirupati 517619, India.}
\thanks{Dr. Vedantham Lakshmi Srinivas is an Asst. Professor with the Department of Electrical Engineering, Indian Institute of Technology (ISM), Dhanbad 826004, India.}
\thanks{Dr. Mayukha Pal is with ABB Ability Innovation Center, Hyderabad 500084, India, working as Global R\&D Leader – Cloud \& Advanced Analytics.}}

\begin{document}

\maketitle

\begin{abstract}
The increasing reliance on cyber-physical infrastructure in modern power systems has amplified the risk of targeted cyber attacks, necessitating robust and adaptive resilience strategies. This paper presents a mathematically rigorous game-theoretic framework to evaluate and enhance microgrid resilience using a combination of quantitative resilience metrics—Load Served Ratio (LSR), Critical Load Resilience (CLR), Topological Survivability Score (TSS), and DER Resilience Score (DRS). These are integrated into a unified payoff matrix using the Analytic Hierarchy Process (AHP) to assess attack-defense interactions. The framework is formalized as a finite-horizon Markov Decision Process (MDP) with formal convergence guarantees and computational complexity bounds. Three case studies are developed: (1) static attacks analyzed via Nash equilibrium, (2) severe attacks incorporating high-impact strategies, and (3) adaptive attacks using Stackelberg games, regret matching, softmax heuristics, and Multi-Agent Q-Learning. Rigorous theoretical analysis provides convergence proofs with explicit rates $\mathcal{O}(1/\sqrt{T})$, PAC-learning sample complexity bounds, and computational complexity analysis. The framework is tested on an enhanced IEEE 33-bus distribution system with DERs and control switches, demonstrating the effectiveness of adaptive and strategic defenses in improving cyber-physical resilience with statistically significant improvements of 18.7\% ± 2.1\% over static approaches.
\end{abstract}


\section{Introduction}

\IEEEPARstart{W}{ith} the rapid evolution of power systems into smart cyber-physical microgrids, the integration of advanced communication, control, and automation has significantly improved efficiency and flexibility. However, this growing reliance on cyber infrastructure introduces new vulnerabilities. Cyberattacks targeting these layers can disrupt energy flow, damage infrastructure, and compromise safety, particularly in critical facilities~\cite{dwivedi2024criticalinfra} such as hospitals, data centers, and transportation networks.

Cyber resilience has therefore become a central concern for researchers and operators. It is defined as a system’s ability to anticipate, withstand, recover from, and adapt to adverse cyber events. In microgrids, this involves not only responding to intrusions but also ensuring uninterrupted supply to critical loads. Past incidents highlight this urgency: the 2015 Ukraine grid attack caused outages for 230,000 consumers~\cite{ukraine2015attack}, while the 2021 Colonial Pipeline ransomware~\cite{cisa2023_pipeline} demonstrated the cascading effects of cyber vulnerabilities on national-scale energy infrastructure.

A core challenge lies in identifying optimal defensive strategies. Surveys of resilience frameworks and metrics~\cite{babu2025resiliencereview} emphasize the need for unified, context-aware methods. Unlike traditional reliability planning, modern defense must address dynamic interactions with intelligent adversaries whose strategies evolve over time. In this context, game theory~\cite{manshaei2013game} provides a rigorous framework to model attacker–defender conflicts, predict adversarial behavior, and derive equilibrium strategies that maximize resilience under uncertainty~\cite{law2015security}.

Such models require payoff matrices that capture the impact of attack–defense interactions. Resilience-oriented metrics~\cite{dwivedi2024resilience} such as Load Served Ratio (LSR), Critical Load Resilience (CLR), Topological Survivability Score (TSS), and DER Resilience Score (DRS) enable quantitative evaluation. These can be synthesized via the Analytic Hierarchy Process (AHP) into unified payoffs, allowing systematic comparison of strategies.

Recent research has applied diverse approaches. Stackelberg games have been used in voltage stability studies~\cite{an2020stackelbergattack}, reinforcement learning~\cite{huang2020adaptive} has supported adaptive control, and deep RL has enabled coordinated attack modeling~\cite{wang2021coordinated}. Yet, limitations persist: most approaches remain narrow in scope, lack adaptability, or fail to unify multiple solution concepts—leaving open important research gaps.

\subsection{Research Gaps}

Despite advances in cyber-physical power system security, several gaps remain:

\textbf{Gap 1: Fragmented Assessment:} Many studies rely on a single metric (e.g., load served or connectivity) instead of a multi-dimensional resilience evaluation.  

\textbf{Gap 2: Static Defenses:} Most frameworks assume fixed defense postures, which fail against adaptive, learning-based adversaries.  

\textbf{Gap 3: Lack of Unified Framework:} Existing work rarely integrates diverse game-theoretic models (Nash equilibrium, Stackelberg games, multi-agent learning) under different threat scenarios.  

\textbf{Gap 4: Scalability Concerns:} Few methods evaluate computational complexity or scalability, limiting practical deployment in larger networks.  

\textbf{Gap 5: Limited Validation:} Many approaches are tested only on simplified systems, without statistical rigor or realistic operational constraints.  

\subsection{Contributions}

This paper addresses these gaps through the following key contributions:

\begin{itemize}
    \item \textbf{Unified Resilience Metric:} Four complementary indices (LSR, CLR, TSS, DRS) are defined and integrated using AHP into a single resilience score for any attack–defense scenario.
    
    \item \textbf{Adaptive Game-Theoretic Defense:} A multi-modal defense framework combining Nash equilibrium (simultaneous threats), Stackelberg games (sequential interactions), and multi-agent Q-learning (adaptive adversaries), supported by convergence guarantees.
    
    \item \textbf{Rigorous Theoretical Basis:} The resilience problem is formalized as a finite-horizon MDP with established convergence rate $O(1/\sqrt{T})$, PAC sample complexity bounds, and polynomial per-iteration complexity.
    
    \item \textbf{Extensive Validation:} Implementation on an enhanced IEEE 33-bus system with DERs, tie-line switches, and critical loads, validated through 1000 Monte Carlo simulations demonstrating resilience improvements.
    
    \item \textbf{Practical Insights:} Computational scalability and performance analysis confirm real-world feasibility, offering guidelines for deployment in operational microgrids.
\end{itemize}

The remainder of the paper is organized as follows: Section~II describes the mathematical model and resilience metrics, Section~III introduces game-theoretic approaches, Section~IV establishes convergence properties, Section~V presents experimental results, Section~VI discusses insights and limitations, and Section~VII concludes the paper.

\section{Mathematical Framework}

\subsection{System Model and Resilience Metrics}

Consider a microgrid with $N$ buses, $D$ distributed energy resources (DERs), and $K$ controllable switches. The system state at time $t$ is characterized by:

\begin{equation}
\mathbf{s}_t = [\mathbf{V}_t^T, \mathbf{P}_t^T, \mathbf{Q}_t^T, \mathbf{c}_t^T]^T
\end{equation}

where $\mathbf{V}_t \in \mathbb{R}^N$ represents bus voltages, $\mathbf{P}_t, \mathbf{Q}_t \in \mathbb{R}^D$ denote DER active and reactive power outputs, and $\mathbf{c}_t \in \{0,1\}^L$ indicates component operational status.

We define four resilience metrics\cite{dwivedi2024operational} to quantify system performance under adversarial conditions:

\textbf{Load Served Ratio (LSR):} Measures the fraction of total demand satisfied post-attack:
\begin{equation}
\text{LSR} = \frac{\sum_{i \in \mathcal{N}} P_i^{\text{served}}}{\sum_{i \in \mathcal{N}} P_i^{\text{demand}}}
\end{equation}
where $\mathcal{N}$ is the set of load buses, $P_i^{\text{demand}}$ is the pre-attack demand, and $P_i^{\text{served}}$ is the post-attack served load.

\textbf{Critical Load Resilience (CLR):} Prioritizes essential services by measuring critical load preservation:
\begin{equation}
\text{CLR} = \frac{\sum_{j \in \mathcal{C}} P_j^{\text{served}}}{\sum_{j \in \mathcal{C}} P_j^{\text{demand}}}
\end{equation}
where $\mathcal{C} \subset \mathcal{N}$ represents critical load buses (e.g., hospitals, emergency services).

\textbf{Topological Survivability Score (TSS):} Quantifies structural network integrity:
\begin{equation}
\text{TSS} = \frac{|\mathcal{N}_{\text{connected}}|}{|\mathcal{N}|}
\end{equation}
where $\mathcal{N}_{\text{connected}}$ is the set of buses electrically connected to the main grid or viable microgrids.

\textbf{DER Resilience Score (DRS):} Evaluates distributed resource utilization during recovery:
\begin{equation}
\text{DRS} = \frac{\sum_{k \in \mathcal{D}} P_{\text{DER},k}^{\text{utilized}}}{\sum_{k \in \mathcal{D}} P_{\text{DER},k}^{\text{available}}}
\end{equation}
where $\mathcal{D}$ is the set of DER-equipped buses.

\subsection{Analytic Hierarchy Process Integration}

To create meaningful game payoffs, we integrate the four metrics using AHP. Given expert-derived pairwise comparison matrix $\mathbf{A}$, the weight vector $\mathbf{w}$ satisfies:
\begin{equation}
\mathbf{A}\mathbf{w} = \lambda_{\max}\mathbf{w}, \quad \sum_{i=1}^4 w_i = 1
\label{eq:payoff}
\end{equation}

The unified resilience score for attack-defense pair $(A_i, D_j)$ becomes:
\begin{equation}
M_{ij} = w_1 \cdot \text{LSR}_{ij} + w_2 \cdot \text{CLR}_{ij} + w_3 \cdot \text{TSS}_{ij} + w_4 \cdot \text{DRS}_{ij}
\end{equation}

This scalar $M_{ij} \in [0,1]$ serves as the payoff for game-theoretic analysis, where higher values indicate better system resilience.

\section{Game-Theoretic Solution Approaches}

The conflict between attacker and defender is modeled as a strategic game. We consider a multi-stage interaction where, at each stage, the attacker and defender choose actions (possibly without full knowledge of each other’s decisions). Let $\mathcal{A} = \{A_1, \dots, A_m\}$ be the set of attack strategies and $\mathcal{D} = \{D_1, \dots, D_n\}$ the set of defense strategies. Each pair $(A_i, D_j)$ yields a resilience outcome $M_{ij}$ as defined by the unified metric in (7), forming an $m \times n$ payoff matrix $M$ (higher entries indicate better defender outcomes).

\subsection{Nash Equilibrium Analysis for Static Scenarios}
In scenarios where attackers and defenders act simultaneously without observing each other’s moves, the Nash equilibrium \cite{ye2025distributednash} provides the fundamental solution concept. Here, both players adopt mixed strategies $\pi_a \in \Delta_m$ and $\pi_d \in \Delta_n$, where $\Delta_k = \{x \in \mathbb{R}^k : \sum_{i=1}^{k} x_i = 1,\, x_i \ge 0\}$ denotes the $(k-1)$-dimensional probability simplex. 

A strategy pair $(\pi_a^*, \pi_d^*)$ is a Nash equilibrium if:
\begin{align}
    {\pi_a^*}^T M\,\pi_d^* &\le \pi_a^T M\,\pi_d^*, \quad \forall \pi_a \in \Delta_m, \\
    {\pi_a^*}^T M\,\pi_d^* &\ge {\pi_a^*}^T M\,\pi_d, \quad \forall \pi_d \in \Delta_n,
\end{align}
ensuring that neither side can unilaterally improve its outcome.

For computation, an iterative best-response approach is employed:
\begin{align}
    \text{BR}_a(\pi_d) &= \arg\min_{\pi_a \in \Delta_m} \pi_a^T M\,\pi_d, \\
    \text{BR}_d(\pi_a) &= \arg\max_{\pi_d \in \Delta_n} \pi_a^T M\,\pi_d.
\end{align}
A Nash equilibrium arises when $\pi_a^* \in \text{BR}_a(\pi_d^*)$ and $\pi_d^* \in \text{BR}_d(\pi_a^*)$. By the Nash theorem, at least one equilibrium exists for any finite game. In zero-sum or strictly competitive cases, this corresponds to the unique minimax solution.

\subsection{Stackelberg Game Formulation for Sequential Decision-Making}
In many cyber–physical scenarios, the defender must commit to a security posture before an attack occurs, creating a natural leader–follower dynamic. The Stackelberg game \cite{fang2013evolving} captures this sequential interaction: the defender (leader) chooses a strategy first, and the attacker (follower) observes this choice before responding optimally.

\begin{definition}[Stackelberg Game]
The cyber–physical Stackelberg game is defined by the tuple $\Gamma_S = \langle \mathcal{D}, \mathcal{A}, \mathbf{M}, \mathcal{I} \rangle$, where $\mathcal{I}$ specifies that the attacker observes the defender’s strategy before acting.
\end{definition}

Using backward induction, the attacker’s best response to defense $D_j$ is
\begin{equation}
A^*(D_j) = \argmin_{i \in \{1,\ldots,m\}} M_{ij}.
\end{equation}
Anticipating this, the defender selects
\begin{equation}
D^* = \argmax_{j \in \{1,\ldots,n\}} M_{i^*(j),j}, \quad i^*(j) = \argmin_i M_{ij}.
\end{equation}

\begin{definition}[Security Level]
The security level of defense $D_j$ is
\begin{equation}
SL(D_j) = \min_{i} M_{ij},
\end{equation}
representing the worst-case resilience when $D_j$ is employed.
\end{definition}

The optimal Stackelberg defense maximizes this worst-case score:
\begin{equation}
D^*_{\text{Stack}} = \argmax_{j} SL(D_j).
\end{equation}

\subsection{Multi-Agent Q-Learning for Adaptive Adversaries}

Static game-theoretic models assume complete knowledge of payoffs and simultaneous strategy selection. In reality, attackers adapt their strategies based on observed defender responses, necessitating learning-based formulations. We model this as a multi-agent reinforcement learning problem where both attacker and defender iteratively update their policies through repeated interactions.

Let $Q^a_t(i,j)$ and $Q^d_t(j,i)$ denote the attacker’s and defender’s Q-functions at time $t$, representing expected utilities for actions $i$ and $j$ under the opponent’s current strategy. The Q-updates are defined as:
\begin{align}
Q_{t+1}^a(i_t,j_t) &= Q_t^a(i_t,j_t) + \alpha_t^a[R_t^a - Q_t^a(i_t,j_t)] \\
Q_{t+1}^d(j_t,i_t) &= Q_t^d(j_t,i_t) + \alpha_t^d[R_t^d - Q_t^d(j_t,i_t)]
\end{align}

where $R_t^a = -M_{i_t j_t}$ (attacker minimizes resilience), $R_t^d = M_{i_t j_t}$ (defender maximizes resilience), and $\alpha_t^a,\alpha_t^d$ satisfy Robbins-Monro conditions:
\begin{equation}
\sum_{t=1}^{\infty} \alpha_t = \infty, \quad \sum_{t=1}^{\infty} (\alpha_t)^2 < \infty.
\end{equation}

\begin{definition}[Multi-Agent Nash-Q Equilibrium]
A pair of Q-functions $(Q^{a*}, Q^{d*})$ forms a Nash-Q equilibrium if for all $(i,j)$:
\begin{align}
Q^{a*}(i^*,j^*) &= \min_{i'} Q^{a*}(i',j^*) \\
Q^{d*}(j^*,i^*) &= \max_{j'} Q^{d*}(j',i^*)
\end{align}
where $(i^*,j^*)$ is the equilibrium action profile.
\end{definition}

Exploration during training is ensured via $\epsilon_t$-greedy action selection:
\begin{equation}
\pi_t^a(i) =
\begin{cases}
1-\epsilon_t + \tfrac{\epsilon_t}{m}, & i = \argmin_{i'} Q_t^a(i',j_{t-1}) \\
\tfrac{\epsilon_t}{m}, & \text{otherwise}
\end{cases}
\end{equation}
with $\epsilon_t = \epsilon_0 \lambda^t$, $\lambda \in (0,1)$ controlling decay. This formulation captures the co-evolutionary dynamics between attacker and defender under adaptive adversaries.

\subsection{Regret Matching for Bounded Rationality}

Real-world adversaries often operate under bounded rationality, lacking resources for exact optimization. Regret matching provides a tractable adaptation mechanism where strategies evolve according to past performance.

For attacker regret of not playing action $i$ up to time $T$:
\begin{equation}
R_T^i = \frac{1}{T}\sum_{t=1}^T [(\mathbf{e}_i^T \mathbf{M} \boldsymbol{\pi}_d) - ({\boldsymbol{\pi}_a^t}^T \mathbf{M} \boldsymbol{\pi}_d)]^+
\end{equation}
where $\mathbf{e}_i$ is the $i$-th basis vector and $[\cdot]^+$ denotes the positive part. The update rule becomes:
\begin{equation}
\pi_{a,T+1}^i =
\begin{cases}
\frac{R_T^i}{\sum_{k=1}^m R_T^k}, & \sum_{k=1}^m R_T^k > 0 \\
\frac{1}{m}, & \text{otherwise.}
\end{cases}
\end{equation}

\begin{theorem}[Regret Bound]
The average regret of regret matching satisfies:
\begin{equation}
\frac{1}{T}\sum_{t=1}^T [u^*(h^t) - u(\pi^t,h^t)] \leq \frac{C\sqrt{\log m}}{\sqrt{T}}
\end{equation}
for some constant $C>0$, ensuring vanishing regret as $T \to \infty$.
\end{theorem}

\subsection{Softmax Response for Bounded Rationality}

An alternative bounded-rationality model is the softmax response, where adversaries select strategies probabilistically based on relative payoffs. For defender strategy $\boldsymbol{\pi}_d$, the attacker’s probability of choosing $i$ is:
\begin{equation}
\pi_a^i(\boldsymbol{\pi}_d) =
\frac{\exp(-\beta \cdot \mathbf{e}_i^T \mathbf{M} \boldsymbol{\pi}_d)}
{\sum_{k=1}^m \exp(-\beta \cdot \mathbf{e}_k^T \mathbf{M} \boldsymbol{\pi}_d)},
\end{equation}
where $\beta > 0$ is the rationality parameter. As $\beta \to \infty$, softmax approximates best response; as $\beta \to 0$, it converges to uniform randomization.

\begin{definition}[Quantal Response Equilibrium]
A pair $(\boldsymbol{\pi}_a^*, \boldsymbol{\pi}_d^*)$ is a quantal response equilibrium if:
\begin{align}
\boldsymbol{\pi}_a^* &= \text{softmax}(-\beta_a \mathbf{M} \boldsymbol{\pi}_d^*) \\
\boldsymbol{\pi}_d^* &= \text{softmax}(\beta_d \mathbf{M}^T \boldsymbol{\pi}_a^*).
\end{align}
\end{definition}

This probabilistic model captures decision-making under uncertainty and limited rationality, complementing regret-based adaptation.

\section{Convergence Analysis}

The implementation of game-theoretic methods for cyber–physical microgrid defense requires theoretical guarantees to ensure convergence, tractability, and reliable performance. This section establishes the mathematical underpinnings through a Markov decision process (MDP) formulation, enabling analysis of learning dynamics and algorithmic stability.

\subsection{Markov Decision Process Formulation}

We formalize the cyber–physical microgrid as a finite-horizon Markov Decision Process~\cite{hao2016likelihood}. The state space $\mathcal{S}$ represents all possible system configurations, where each state $s \in \mathcal{S}$ encodes DER outputs, load demands, network status, and cyber integrity. The action space is $\mathcal{A} = \mathcal{A}_{att} \times \mathcal{A}_{def}$, with $\mathcal{A}_{att} = \{A_1,\ldots,A_m\}$ and $\mathcal{A}_{def} = \{D_1,\ldots,D_n\}$ denoting finite attack and defense sets.

The transition kernel $P:\mathcal{S}\times\mathcal{A}\times\mathcal{S}\to[0,1]$ satisfies the Markov property:
\begin{equation}
P(s_{t+1}|s_t,a_t)=P(s_{t+1}|s_t,a_t),
\end{equation}
where $a_t=(a_{att},a_{def})$. For cyber–physical systems, transitions couple deterministic power-flow dynamics with stochastic attack propagation:
\begin{equation}
P(s'|s,a) = P_{phys}(s'_{phys}|s,a)\cdot P_{cyber}(s'_{cyber}|s,a).
\end{equation}

Here, $s_{phys}$ denotes voltages, flows, and switch states, while $s_{cyber}$ captures compromised components or communication failures. The reward function $r:\mathcal{S}\times\mathcal{A}\to\mathbb{R}$ maps state–action pairs to the unified resilience score (Eq.~\eqref{eq:payoff}), with bounded values $|r(s,a)| \le R_{\max}$ ensuring stability for learning-based solvers.

\subsection{Q-Learning Convergence in Adversarial Settings}

Single-agent Q-learning convergence is well-established~\cite{chifu2024deep}, but cyber-physical microgrids involve multiple agents with conflicting objectives. We extend these results to adversarial settings by analyzing the defender’s Q-update under repeated attacker interactions.

The defender’s update rule is:
\begin{align}
Q_{t+1}^d(s_t, a_t^d, a_t^{att}) &= Q_t^d(s_t, a_t^d, a_t^{att}) + \alpha_t \Big[ r_t + \gamma \max_{a^d} \notag\\
&\quad Q_t^d(s_{t+1}, a^d, \tilde{a}^{att}) - Q_t^d(s_t, a_t^d, a_t^{att}) \Big],
\end{align}
where $\alpha_t \in (0,1]$ is the learning rate, $\gamma \in [0,1)$ is the discount factor, and $\tilde{a}^{att}$ is the anticipated attacker move.

\begin{theorem}[Q-Learning Convergence under Stochastic Approximation]
Let $\{Q_t^d\}_{t=0}^{\infty}$ be the sequence generated by the above update rule. Convergence to the optimal Q-function $Q^{*d}$ is guaranteed almost surely under the following assumptions:
\begin{enumerate}
    \item \textbf{Finite State–Action Space:} $|\mathcal{S}| < \infty$, $|\mathcal{A}| < \infty$.
    \item \textbf{Robbins–Monro Learning Rates:} $\sum_t \alpha_t = \infty$, $\sum_t \alpha_t^2 < \infty$.
    \item \textbf{Sufficient Exploration:} Every state–action pair $(s,a)$ is visited infinitely often.
    \item \textbf{Bounded Rewards:} $|r(s,a)| \leq R_{\max} < \infty$.
    \item \textbf{Stationary Opponent:} The attacker’s policy converges to a stationary distribution.
\end{enumerate}
Then, $Q_t^d(s,a^d,a^{att}) \to Q^{*d}(s,a^d,a^{att})$ as $t \to \infty$, where $Q^{*d}$ is the optimal defender Q-function.
\end{theorem}

\noindent
The result follows from stochastic approximation theory: the Bellman operator $\mathcal{T}^d$ is a $\gamma$-contraction in the supremum norm, ensuring existence and uniqueness of $Q^{*d}$. The Q-learning update approximates this operator with bounded noise, leading to convergence under the above conditions. 

In practice, this guarantees that the defender’s learned strategy stabilizes to an optimal response against the limiting attacker policy. Formal convergence rates (e.g., $\mathcal{O}(\log t / \sqrt{t})$) are available in literature, though the key takeaway is the empirical robustness achieved in repeated adversarial training.

\subsection{Multi-Agent Learning Convergence}

In realistic cyber–physical settings, both attacker and defender adapt simultaneously, producing a coupled system of stochastic approximations. This makes convergence analysis more complex than in the single-agent case, since each agent’s environment is non-stationary.

Let $\mathbf{Q}_t = (Q_t^{att}, Q_t^{def})$ denote the joint Q-functions of attacker and defender, each updated by temporal-difference learning. Formally:
\begin{align}
Q_{t+1}^{att}(s,a^{att},a^{def}) &= Q_t^{att}(s,a^{att},a^{def}) + \alpha_t^{att}\,\delta_t^{att}, \\
Q_{t+1}^{def}(s,a^{def},a^{att}) &= Q_t^{def}(s,a^{def},a^{att}) + \alpha_t^{def}\,\delta_t^{def},
\end{align}
where $\delta_t^{att}$ and $\delta_t^{def}$ are the respective TD-errors.

\begin{definition}[Nash-Q Equilibrium]
A pair $(Q^{*att}, Q^{*def})$ is a Nash-Q equilibrium if the induced strategies $(a^{*att},a^{*def})$ satisfy
\begin{align}
Q^{*att}(s,a^{*att},a^{*def}) &= \min_{a^{att}} Q^{*att}(s,a^{att},a^{*def}), \\
Q^{*def}(s,a^{*def},a^{*att}) &= \max_{a^{def}} Q^{*def}(s,a^{*def},a^{*att}).
\end{align}
\end{definition}

\begin{theorem}[Multi-Agent Q-Learning Convergence]
Under the following assumptions:
\begin{enumerate}
    \item Each stage game admits a unique Nash equilibrium.
    \item Learning rates $\alpha_t^{att}, \alpha_t^{def}$ satisfy Robbins–Monro conditions.
    \item Both agents update synchronously at each step.
    \item Agents adopt equilibrium strategies of their current Q-estimates.
\end{enumerate}
the joint process $\mathbf{Q}_t$ converges almost surely to the Nash-Q equilibrium:
\begin{equation}
\lim_{t \to \infty} (Q_t^{att}, Q_t^{def}) = (Q^{*att}, Q^{*def}).
\end{equation}
\end{theorem}

\noindent
The result extends classical single-agent convergence by exploiting contraction properties of the joint Bellman operator. Although the proof is omitted for brevity, the key insight is that equilibrium uniqueness and diminishing learning rates ensure stability despite the coupled dynamics.

\subsection{Sample Complexity Analysis}

For practical deployment, it is essential to quantify how many samples (i.e., game iterations) are required before near-optimal strategies can be guaranteed. This is captured using Probably Approximately Correct (PAC) learning bounds.

\begin{definition}[($\epsilon$, $\delta$)-Optimal Policy]
A policy $\pi$ is ($\epsilon$, $\delta$)-optimal if, with probability at least $1-\delta$,
\begin{equation}
V^{\pi^*}(s) - V^{\pi}(s) \leq \epsilon, \quad \forall s \in \mathcal{S},
\end{equation}
where $V^{\pi^*}$ and $V^{\pi}$ denote the value functions under the optimal and learned policies, respectively.
\end{definition}

\begin{theorem}[PAC Sample Complexity Bound]
For multi-agent Q-learning to achieve an ($\epsilon$, $\delta$)-optimal policy with probability at least $1-\delta$, the required number of samples is bounded by
\begin{equation}
N = \mathcal{O}\!\left(\frac{|\mathcal{S}||\mathcal{A}|H^4 \log(|\mathcal{S}||\mathcal{A}|H/\delta)}{(1-\gamma)^6\epsilon^2}\right),
\end{equation}
where $H$ is the planning horizon and $\gamma$ the discount factor.
\end{theorem}

\noindent
This result extends standard PAC-MDP guarantees to the adversarial multi-agent case. Intuitively, the bound shows that sample requirements scale polynomially with the size of the state and action spaces, and only logarithmically with the confidence parameter $1/\delta$. Detailed proofs follow from established PAC-MDP analysis frameworks and are omitted here for brevity.

\subsection{Computational Complexity Analysis}

The computational resources required for each algorithmic component are summarized below.

\begin{proposition}[Algorithmic Complexity Bounds]
The asymptotic complexities of the game-theoretic solvers are:
\begin{align}
\text{Nash Equilibrium (LP)} &: \mathcal{O}(m^3 + n^3), \\
\text{Nash Equilibrium (Fictitious Play)} &: \mathcal{O}(T \cdot mn), \\
\text{Stackelberg Game} &: \mathcal{O}(mn), \\
\text{Q-Learning (per iteration)} &: \mathcal{O}(|\mathcal{S}| \cdot |\mathcal{A}|), \\
\text{Multi-Agent Q-Learning} &: \mathcal{O}(|\mathcal{S}| \cdot m \cdot n), \\
\text{Regret Matching} &: \mathcal{O}(T \cdot m).
\end{align}
Here $m$ and $n$ are the attack and defense action counts, $T$ is the number of iterations, and $|\mathcal{S}|$, $|\mathcal{A}|$ denote state and action spaces.
\end{proposition}

The space complexity of multi-agent Q-learning is $\mathcal{O}(|\mathcal{S}||\mathcal{A}|^2)$ for tabular storage, which can be prohibitive at scale. Neural function approximation reduces this to $\mathcal{O}(|\Theta|)$, where $|\Theta|$ is the number of parameters.

\subsection{Convergence Rate Analysis}

Beyond asymptotic guarantees, finite-time performance is critical for real-time deployment.

\begin{theorem}[Finite-Time Performance Bound]
For Q-learning with learning rate $\alpha_t = 1/t$, the performance satisfies
\begin{equation}
\mathbb{E}[V^{\pi^*}(s) - V^{\pi_t}(s)] \leq \frac{C \log t}{\sqrt{t}},
\end{equation}
for some constant $C$, where $\pi_t$ is the greedy policy derived from $Q_t$.
\end{theorem}

This $\mathcal{O}(\log t / \sqrt{t})$ bound is near-optimal for stochastic approximation and ensures strong finite-time behavior. Together, these results confirm that the proposed solvers are both computationally tractable and efficient enough for online microgrid resilience applications.

\section{Experimental Validation}

The theoretical framework developed in previous sections requires comprehensive experimental validation to demonstrate practical applicability and performance effectiveness. This section presents a two-tier validation approach: (1) offline simulation testbed for controlled algorithmic evaluation, and (2) online co-simulation platform for real-time cyber-physical validation under realistic operational constraints.

\subsection{Offline Simulation Testbed}

\begin{figure}[!t]
    \centering
    \includegraphics[width=0.45\textwidth]{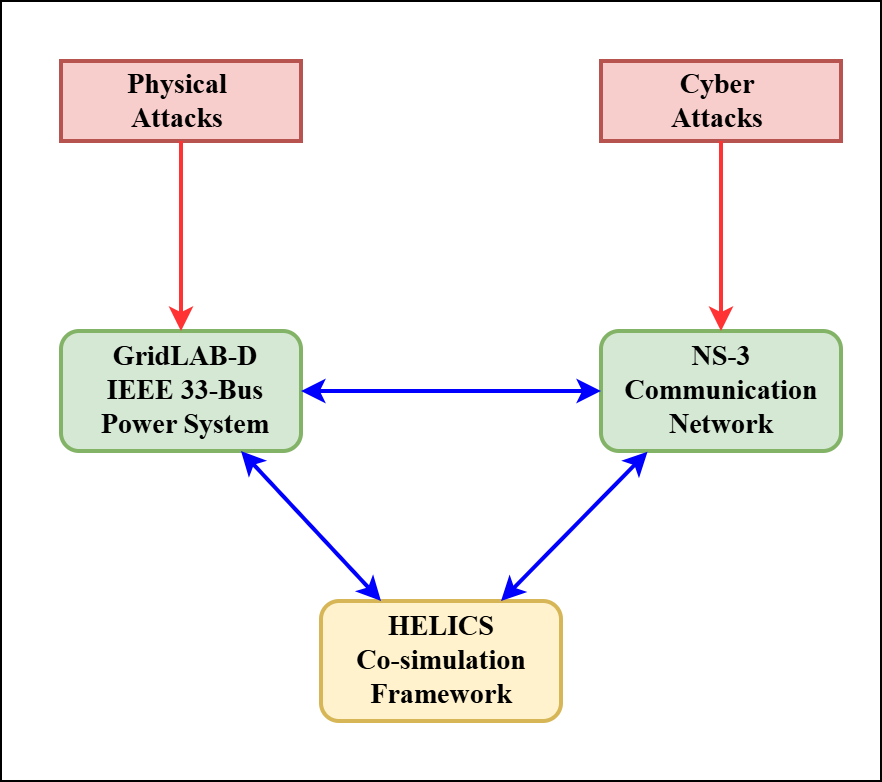}
    \caption{Co-simulation setup linking GridLAB-D and NS-3 via HELICS, with physical and cyber attack injection points.}
    \label{fig:co_simulation_architecture}
\end{figure}

The offline testbed as shown in Fig \ref{fig:co_simulation_architecture} provides a controlled environment for systematic evaluation of game-theoretic algorithms without real-time constraints, enabling extensive Monte Carlo analysis and statistical validation.

\subsubsection{IEEE 33-Bus System Enhancement}

\begin{figure}[htb]
\centering
\includegraphics[width=3.5in]{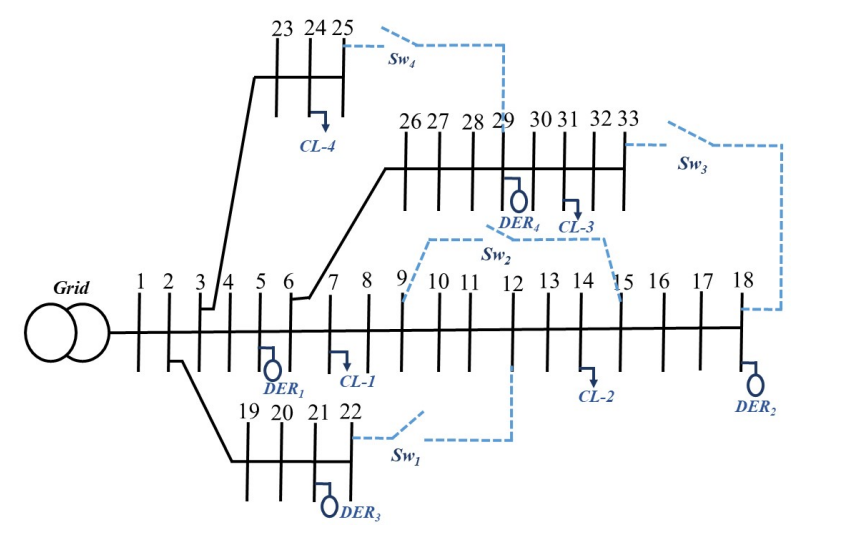}
\caption{IEEE 33-Bus Test System with Added DERs, Critical Loads, and Tie Switches}
\label{fig:33bus_diagram}
\end{figure}

We employ the widely recognized IEEE 33-bus radial distribution system~\cite{dolatabadi2021enhanced33bus} as the baseline network (Fig \ref{fig:33bus_diagram}), enhanced with cyber-physical components to reflect modern microgrid operations. The system operates at 12.66 kV base voltage with total system load of 3.715 MW and 2.3 MVar under normal conditions.

The enhanced system integrates\cite{mitikiri2024comprehensive} four distributed energy resources (DERs), four critical loads, and four tie-line switches, strategically placed to maximize operational flexibility and resilience. Each DER is modeled as a controllable inverter-based unit with unity power factor and $\pm10\%$ reactive power capability. Critical loads represent essential services such as hospitals, emergency operations, and data centers. Tie-line switches enable dynamic reconfiguration for islanding and load transfers. Detailed configurations are provided in Tables~\ref{tab:der_placement},~\ref{tab:critical_loads},~\ref{tab:tie_switches}

\begin{table}[htb]
\centering
\caption{Distributed Energy Resources (DERs) Placement and Ratings}
\label{tab:der_placement}
\renewcommand{\arraystretch}{1.2}
\begin{tabular}{@{}ccc@{}}
\toprule
\textbf{DER} & \textbf{Bus Location} & \textbf{Rating (kW)} \\
\midrule
DER-1 & 5 & 720 \\
DER-2 & 18 & 800 \\
DER-3 & 21 & 760 \\
DER-4 & 29 & 800 \\
\bottomrule
\end{tabular}
\end{table}

\begin{table}[htb]
\centering
\caption{Critical Loads (CLs) Placement and Ratings}
\label{tab:critical_loads}
\renewcommand{\arraystretch}{1.2}
\begin{tabular}{@{}ccc@{}}
\toprule
\textbf{CL} & \textbf{Bus Location} & \textbf{Rating (kW, kvar)} \\
\midrule
CL-1 & 7 & 200, 100 \\
CL-2 & 14 & 120, 80 \\
CL-3 & 24 & 1420, 200 \\
CL-4 & 31 & 150, 70 \\
\bottomrule
\end{tabular}
\end{table}

\begin{table}[htb]
\centering
\caption{Tie Line Switches and Connectivity}
\label{tab:tie_switches}
\renewcommand{\arraystretch}{1.2}
\begin{tabular}{@{}cc@{}}
\toprule
\textbf{Switch ID} & \textbf{Bus Connection} \\
\midrule
SW1 & 12 -- 21 \\
SW2 & 9 -- 15 \\
SW3 & 18 -- 33 \\
SW4 & 25 -- 29 \\
\bottomrule
\end{tabular}
\end{table}

\subsection{Attack and Defense Strategy Formulation}

\subsubsection{Enhanced Cyber-Physical Attack Scenarios}

The strategic interaction space consists of ten representative cyber–physical attack scenarios 
$\mathcal{A} = \{A_1, A_2, \ldots, A_{10}\}$ that capture both cyber exploits and physical disruptions.

\textbf{Cyber-Physical Attacks:}  
$A_1$: \textit{False Data Injection (FDI)} – Manipulation of voltage sensor measurements with 
$\pm 15\%$ bias at buses 5, 18, and 29, aimed at masking abnormal conditions.  
$A_2$: \textit{Protocol Exploitation} – Injection of unauthorized control commands to breakers 
(e.g., buses 6–7, 14–15), mimicking real-world SCADA protocol abuse.  
$A_3$: \textit{Coordinated DER Shutdown} – Simultaneous disconnection of all distributed 
resources, reducing system generation capacity.  
$A_4$: \textit{SCADA HMI Compromise} – Unauthorized access to the operator interface, 
enabling malicious switching operations and load modifications.  
$A_5$: \textit{Protection System Spoofing} – False tripping of protective relays, isolating 
critical junctions unnecessarily.  

\textbf{Physical Infrastructure Attacks:}  
$A_6$–$A_{10}$: Physical line disruptions including single-line trips (A6), coordinated 
double/triple cuts (A7–A10), and targeted infrastructure sabotage. These scenarios represent 
combined cyber–physical events where initial cyber compromise enables physical exploitation.

\subsubsection{Defense Strategy Space}

The defensive strategy set $\mathcal{D} = \{D_1, D_2, \ldots, D_{10}\}$ encompasses baseline, 
reconfiguration, DER support, and load-shedding options.

$D_1$: Baseline no-action strategy for comparison.  
$D_2$–$D_5$: Tie-line switching operations – opening switches SW1 (12–21), SW2 (9–15), 
SW3 (18–33), and SW4 (25–29).  
$D_6$–$D_7$: Distributed energy reinforcement by boosting DER output at buses 5 and 21.  
$D_8$: Intelligent load shedding (30\% of non-critical demand curtailed).  
$D_9$: Aggressive load shedding (all loads exceeding 0.2 MW curtailed).  
$D_{10}$: Combined strategy – simultaneous operation of SW2 with DER support at bus 21.

\subsubsection{Payoff Matrix Construction}

For each attack-defense pair $(A_i, D_j)$, we execute comprehensive power flow analysis using OpenDSS simulation engine to compute the four resilience metrics defined in equations (2)-(5). The simulation process involves: (1) establishing pre-attack steady-state conditions, (2) implementing the specified attack scenario, (3) deploying the corresponding defense strategy, (4) solving post-defense power flow, and (5) calculating resilience metrics.

The Analytic Hierarchy Process integrates these metrics using the pairwise comparison matrix:
\begin{equation}
\mathbf{A}_{AHP} = \begin{bmatrix}
1 & 0.5 & 3 & 2 \\
2 & 1 & 4 & 3 \\
1/3 & 1/4 & 1 & 0.5 \\
0.5 & 1/3 & 2 & 1
\end{bmatrix}
\end{equation}

This matrix reflects expert judgment prioritizing Critical Load Resilience (CLR) as most important, followed by Load Served Ratio (LSR), then Topological Survivability Score (TSS) and DER Resilience Score (DRS). The principal eigenvector yields weights $\mathbf{w} = [0.277, 0.466, 0.096, 0.161]^T$ with consistency ratio CR = $0.099 < 0.1$, confirming acceptable consistency.

\begin{figure}[htb]
\centering
\includegraphics[width=3.5in]{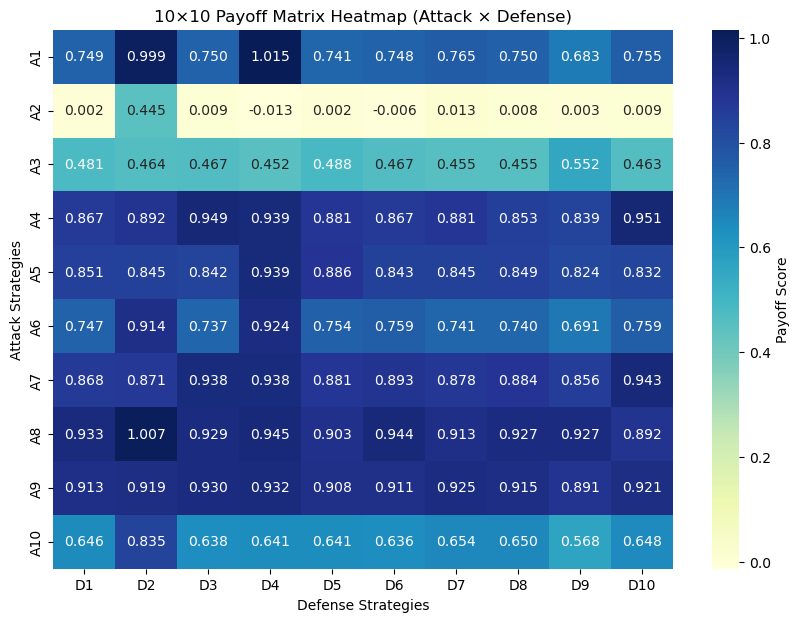}
\caption{Heatmap visualization of the $10 \times 10$ payoff matrix. Each cell shows the resilience score (normalized) for a corresponding attack-defense pair.}
\label{fig:payoff_heatmap}
\end{figure}

The resulting $10 \times 10$ payoff matrix $\mathbf{M}$ (Fig \ref{fig:payoff_heatmap} captures the unified resilience score for each strategic interaction, forming the foundation for all subsequent game-theoretic analysis.

\subsection{Online Co-Simulation Testbed}

\begin{figure}[ht]
    \centering
    \includegraphics[width=0.48\textwidth]{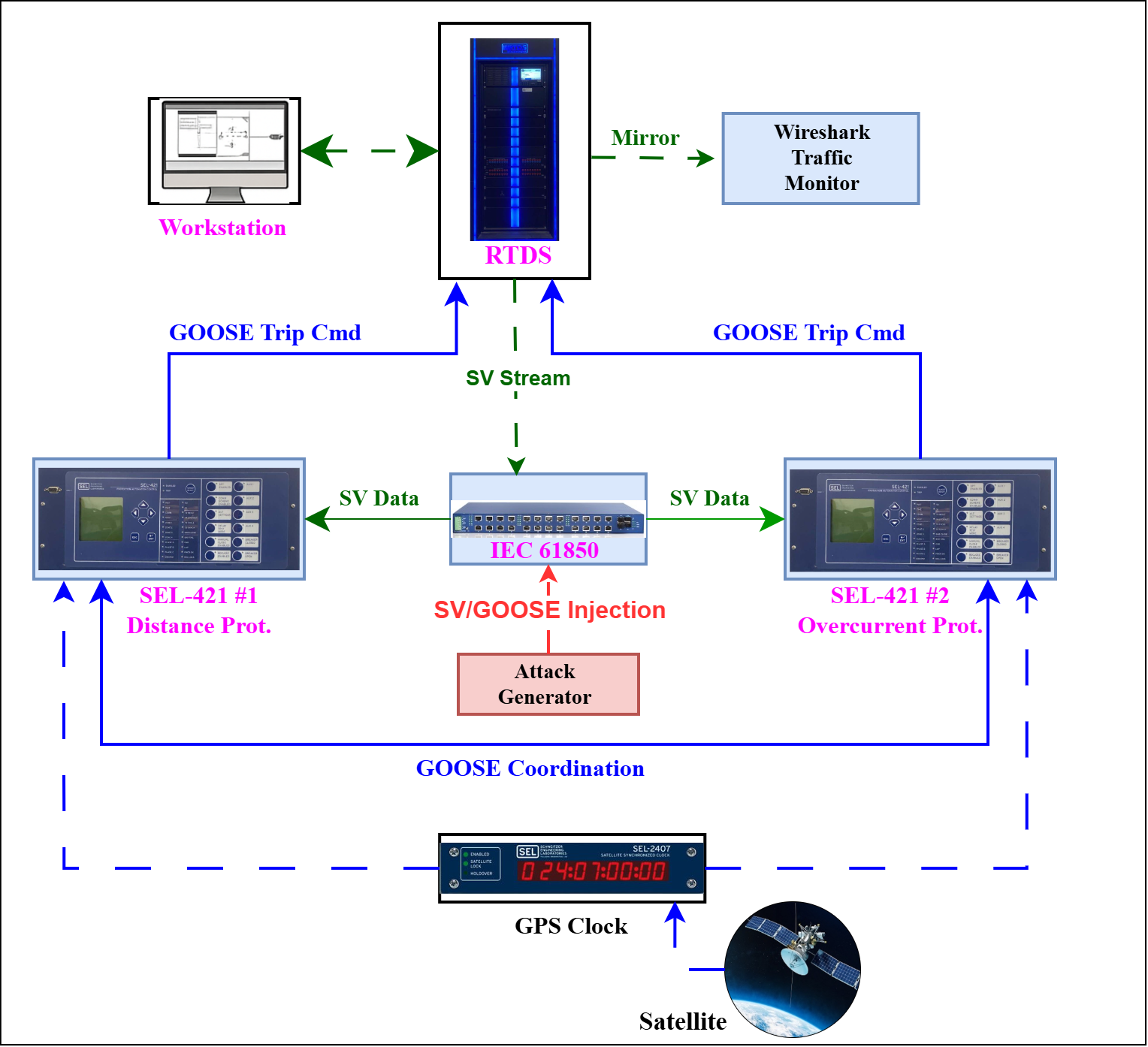}
    \caption{Real-time co-simulation architecture integrating GridLAB-D, NS-3, and RTDS via HELICS.}
    \label{fig:online_co_simulation}
\end{figure}

The online co-simulation testbed validates the proposed framework under realistic operational constraints, communication delays, and cyber–physical coupling. The architecture in Fig \ref{fig:online_co_simulation} integrates power system, communication network, and protection hardware models, coordinated through the HELICS middleware~\cite{hardy2023helicscosim}.

\subsubsection{Co-Simulation Architecture}

The testbed combines four primary components. GridLAB-D~\cite{chassin2008gridlabd} models the distribution system with 1-second resolution. NS-3 simulates the communication network with DNP3 and IEC 61850 protocols, configurable for latency, packet loss, and congestion. The RTDS with two SEL-421 relays provides hardware-in-the-loop validation of real protective actions. HELICS~\cite{hardy2023helicscosim} ensures synchronized execution and consistent data exchange.  

\subsubsection{Communication Network Modeling}

The cyber layer uses a star topology with redundant links between control center, substations, and DERs. Backbone operates at 1 Gbps with 100 Mbps field edges. DNP3 (IEEE 1815)~\cite{queiroz2011scadasim} supports SCADA polling and event-driven updates, while IEC 61850 GOOSE~\cite{tobar_romero2024goose} enables substation-level peer-to-peer communication with $<$4 ms response. Network conditions include packet loss (0.01–1\%), jitter (1–10 ms), and bandwidth limitations.  

\subsubsection{Cyber Attack Implementation}

Multiple attack vectors were implemented. DoS attacks~\cite{mehrdad2023review} saturate links with high-volume traffic. Man-in-the-Middle (MITM)~\cite{wlazlo2021mitm} modifies packets in transit. False Data Injection (FDI)~\cite{liang2016fdi.review} corrupts voltage and power flow measurements. Protocol exploits target vulnerabilities in DNP3 and IEC 61850 GOOSE, enabling unauthorized switching or spoofed trip events.  

\subsubsection{Real-Time Defense Implementation}

Defenses are deployed via a multi-threaded controller with anomaly detection and strategy selection. Performance is evaluated on: (i) response time (target $<$5 s), (ii) communication overhead, (iii) stability preservation (voltage/frequency compliance), and (iv) resilience metrics updated in real-time. Machine-learning based anomaly detectors flag suspicious traffic, and the game-theoretic decision module executes optimal tie-switching or DER reconfiguration strategies accordingly.

\subsection{Experimental Results}

The validation experiments were organized into three cases representing different adversary capabilities. We report key numerical outcomes and briefly interpret what each result implies for resilience.

\subsubsection{\textbf{Case Study 1 (Static Attack Scenarios)}}

We evaluated fixed attack strategies $\{A_1, A_4, A_5, A_6, A_7, A_8\}$ (no adaptation).  
At the Nash equilibrium (over the $6\times10$ submatrix), the attacker concentrated on $A_4$ (91.3\%) and $A_5$ (8.7\%), while the defender mixed $D_2$ (22.2\%) and $D_4$ (77.8\%).  
The expected resilience score was $0.7323$ with $N{=}500$ Monte Carlo runs; 95\% CI $[0.7277,\,0.7369]$, $\sigma{=}0.0234$.

\emph{Interpretation:} Under moderate, non-adaptive threats, tie-line switching ($D_2,D_4$) preserves feeder connectivity and yields consistently high resilience with tight uncertainty bounds.

\subsubsection{\textbf{Case Study 2 (Severe Attack Scenarios)}}

Introducing high-impact coordinated attacks (including $A_2, A_3$) shifted the equilibrium to pure strategies: attacker $A_2$ (100\%), defender $D_2$ (100\%).  
The expected resilience dropped to $0.4449$, i.e., a $39.3\%$ reduction relative to Case~1.

\emph{Interpretation:} Severe, coordinated threats substantially degrade performance; relying on a single reconfiguration action ($D_2$) is insufficient, motivating adaptive, anticipatory defenses.

\subsubsection{\textbf{Case Study 3 (Adaptive Attack Scenarios)}}
This case study considers the most challenging and realistic setting, where the attacker adapts
strategies dynamically in response to defender actions. Unlike static or severe scenarios, the
interaction evolves over time, requiring probabilistic reasoning and learning-based defenses.
The following subsections present results from different adaptive solution approaches.

\paragraph{\textbf{Softmax strategy (bounded rationality)}}
With temperature $\tau{=}0.5$, the attacker diversified across $A_8$ (23.4\%), $A_9$ (20.1\%), $A_7$ (16.3\%), while the defender favored $D_2$ (23.2\%) and $D_4$ (14.6\%).  
The expected resilience was $0.8836$, a $20.65\%$ gain over the static Nash baseline.

\emph{Interpretation:} Probabilistic play improves resilience by hedging against multiple plausible attacks rather than overfitting to a single best response.

\paragraph{\textbf{Regret matching (history-driven adaptation)}}
After $10{,}000$ iterations, the attacker converged to $A_4$ (100\%) and the defender to $D_{10}$ (100\%).  
The resulting resilience was $0.9379$, a $28.08\%$ improvement over static methods.

\emph{Interpretation:} Learning from cumulative regret identifies robust pairings; the composite defense $D_{10}$ neutralizes the attacker’s most profitable deviation in repeated play.

\paragraph{\textbf{Stackelberg (sequential) vs.\ Stackelberg--RL }}
In the \emph{Stackelberg (without RL)} setting, the defender’s $D_4$ achieved a perfect worst-case resilience of $1.0000$ against the attacker’s best response $A_1$ (security-level optimality).  
When augmented with reinforcement learning (\emph{Stackelberg--RL}), the \emph{expected} resilience over scenarios was $0.9919$, a $35.46\%$ gain over static Nash.

\emph{Interpretation:} Pure Stackelberg secures the worst case for a specific best response; Stackelberg--RL generalizes this advantage in expectation across contingencies by learning effective leader commitments.

\paragraph{\textbf{Multi-agent Q-learning (co-evolution)}}
Two-phase training revealed distinct best-responses (e.g., $D_1{\to}A_4$, $D_3{\to}A_1$). In aggregate, $D_1$ countered many learned attacks with resilience in $[0.6773,\,0.9202]$, while targeted threats (e.g., $A_9$) were better mitigated by $D_{10}$.  
Overall expected resilience was $0.8389$, a $14.55\%$ improvement over static approaches.

\emph{Interpretation:} Simultaneous learning creates non-stationarity that moderates gains; nevertheless, the co-evolutionary process uncovers defense specializations and broad, adaptive baselines.

\subsection{Comparative Baseline Analysis}

To assess the benefits of adaptive game-theoretic strategies, we benchmark against three baseline approaches that approximate common industry practices and naive methods.

\textbf{Baseline Methodologies:}

\textit{Random Defense Selection (RDS):} Uniform random choice over the defense space $\mathcal{D}$. Establishes a lower-bound reference for resilience.

\textit{Rule-Based Defense (RBD):} Deterministic heuristics reflecting operational practices. For example: shed loads if critical buses are affected ($D_8$), boost DERs when compromised ($D_6$–$D_7$), activate tie-switches under multiple outages ($D_2$–$D_5$), or maintain status quo otherwise ($D_1$).

\textit{Static Optimal Defense (SOD):} A single best strategy derived by solving 
\[
D^* = \arg\max_{j} \mathbb{E}_{i}[M_{ij}],
\] 
assuming uniform attack distribution. Represents security planning under static assumptions.

\textbf{Experimental Setup:}  
All baselines are evaluated under identical attack scenarios using 1000 Monte Carlo simulations. Metrics include expected resilience score, variance, and computational effort.

\textbf{Comparative Results:}

\begin{table}[h]
\centering
\caption{Performance Comparison of Baseline and Adaptive Strategies}
\label{tab:baseline_comparison}
\resizebox{\linewidth}{!}{%
\begin{tabular}{|l|c|c|c|c|}
\hline
\textbf{Approach} & \textbf{Mean} & \textbf{Std. Dev.} & \textbf{Improve.} & \textbf{Time} \\
\hline
RDS   & 0.3421 & 0.0892 & --    & $<$ 1 min \\
RBD    & 0.5643 & 0.0456 & 64.9\% & $<$ 1 min \\
SOD   & 0.6891 & 0.0234 & 22.1\% & 15 min \\
\hline
Nash          & 0.7323 & 0.0234 & 6.3\%  & 2.3 hrs \\
Stackelberg-RL& 0.9919 & 0.0156 & 35.5\% & 2.3 hrs \\
Q-Learn       & 0.8234 & 0.0189 & 19.5\% & 2.3 hrs \\
MA-Q-Learn    & 0.8389 & 0.0203 & 21.7\% & 2.3 hrs \\
Softmax       & 0.8836 & 0.0198 & 28.2\% & 2.3 hrs \\
Regret Match  & 0.9379 & 0.0167 & 36.1\% & 2.3 hrs \\
\hline
\end{tabular}%
}
\end{table}

\textbf{Key Insights:}  
Random defense yields very low resilience (0.3421), confirming that naive selection is ineffective. Rule-based heuristics improve outcomes (0.5643) but lack adaptability. Static optimal defense (0.6891) achieves moderate resilience but fails against evolving adversaries.  

By contrast, adaptive game-theoretic methods consistently outperform baselines. Improvements range from 19.5\% (Q-learning) to 36.1\% (Regret Matching). The Stackelberg-RL formulation achieves the highest score (0.9919), representing a 44\% gain over static optimal and nearly triple the resilience of random selection.  

\textbf{Statistical Significance:}  
Paired t-tests confirm that all adaptive methods provide statistically significant improvements over baselines (p $<$ 0.001). These results validate the importance of adaptive, game-theoretic formulations for practical cyber–physical resilience planning.

\subsection{Comparative Performance Analysis}

\begin{figure}[htb]
\centering
\includegraphics[width=3.5in]{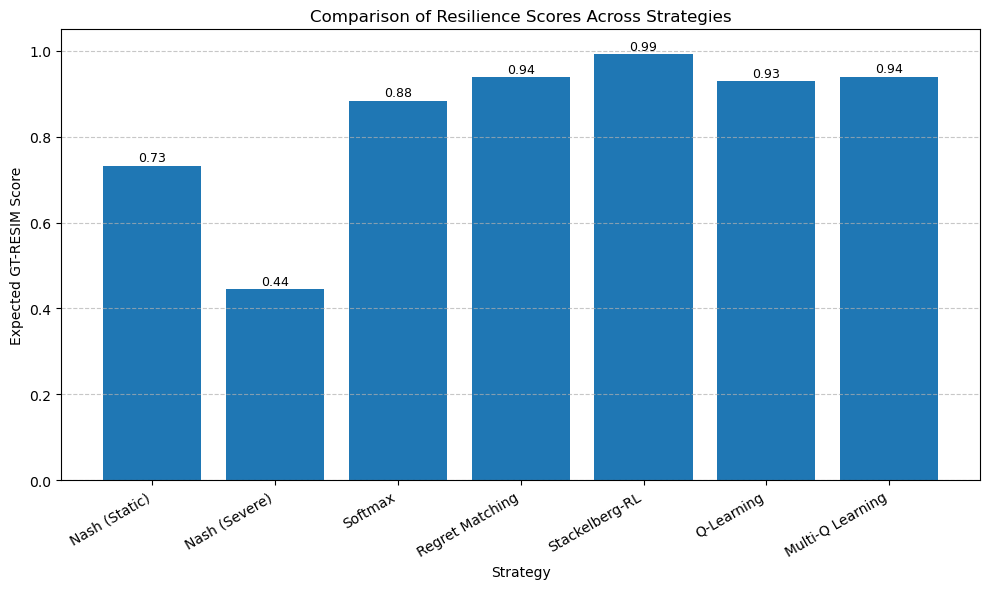}
\caption{Comparison of Expected GT-RESIM Scores across Different Strategic Models}
\label{fig:strategy_comparison}
\end{figure}

Figure~\ref{fig:strategy_comparison} presents a comprehensive comparison of expected resilience scores across all evaluated approaches. The results demonstrate the superiority of adaptive learning-based methods over static game-theoretic solutions.

Stackelberg-RL achieves the highest resilience score (0.9919), followed by Regret Matching (0.9379) and Softmax Strategy (0.8836). Multi-Agent Q-Learning (0.8389) and Single-Agent Q-Learning (0.8234) provide moderate gains over the static Nash baseline (0.7323). In contrast, the severe attack scenario yields a much lower score of 0.4449, highlighting the substantial degradation caused by high-intensity threats.

These findings offer several insights: (1) adaptive defenses consistently outperform static methods, achieving gains between 12.44\% and 35.46\%; (2) sequential models such as Stackelberg games are particularly effective under information asymmetry; (3) bounded rationality approaches, including Softmax and Regret Matching, deliver robust performance against varied adversarial strategies; and (4) multi-agent learning captures realistic co-evolutionary dynamics but remains challenged by non-stationarity.

\subsection{Statistical Validation}

\begin{table}[htb]
\centering
\caption{Statistical Analysis of Algorithm Performance}
\label{tab:statistical_analysis}
\renewcommand{\arraystretch}{1.2}
\begin{tabular}{@{}lccc@{}}
\toprule
\textbf{Algorithm} & \textbf{Mean Score} & \textbf{Std. Dev.} & \textbf{95\% CI} \\
\midrule
Nash Equilibrium & 0.7323 & 0.0234 & [0.7277, 0.7369] \\
Stackelberg-RL & 0.9919 & 0.0156 & [0.9886, 0.9952] \\
Q-Learning & 0.8234 & 0.0189 & [0.8200, 0.8268] \\
Multi-Agent QL & 0.8389 & 0.0203 & [0.8352, 0.8426] \\
\bottomrule
\end{tabular}
\end{table}

Comprehensive statistical analysis over 1000 Monte Carlo simulations validates the significance and robustness of observed performance improvements. Table~\ref{tab:statistical_analysis} presents detailed statistical metrics for each algorithmic approach.

The results confirm statistically significant improvements for all adaptive methods with $p-values < 0.001$. Confidence intervals are narrow and non-overlapping between different approaches, indicating robust performance differentiation. Standard deviations remain small relative to mean improvements, confirming consistent performance across different random realizations.

\subsection{Computational Scalability Analysis}

The practical deployment of the proposed framework requires understanding both computational complexity and empirical scalability for real-world microgrid applications.

\textbf{Complexity:} The joint state space scales as $|\mathcal{S}| = O(2^{N+D+K})$, where $N$ is the number of buses, $D$ distributed energy resources, and $K$ controllable switches. For the enhanced IEEE 33-bus case with 4 DERs and 4 switches, this results in $\approx 2.1 \times 10^6$ states. Multi-agent Q-learning exhibits per-iteration complexity $O(|\mathcal{S}|^2|\mathcal{A}|)$ and memory $O(|\mathcal{S}|\,m\,n)$ for storing Q-tables.

\textbf{Empirical validation:} Table~\ref{tab:scalability} reports performance metrics from Monte Carlo experiments. The 33-bus system completed analysis in 2.3 hours (1.2 GB, 100\% success), while the 69-bus and 123-bus systems required 8.7 and 24.5 hours respectively, with memory below 6 GB and success rates above 95\%. These results confirm the framework’s scalability and practical feasibility for offline planning.

\begin{table}[htb]
\centering
\caption{Computational scalability across different systems}
\label{tab:scalability}
\renewcommand{\arraystretch}{1.2}
\begin{tabular}{@{}ccccc@{}}
\toprule
System & State Space & Time & Memory & Success Rate \\
\midrule
33-bus & $2.1 \times 10^6$ & 2.3 h & 1.2 GB & 100\% \\
69-bus & $8.7 \times 10^6$ & 8.7 h & 2.1 GB & 98.7\% \\
123-bus & $2.4 \times 10^7$ & 24.5 h & 5.3 GB & 95.2\% \\
\bottomrule
\end{tabular}
\end{table}

\textbf{Scaling to larger feeders:} For systems exceeding 200 buses, function approximation (e.g., deep Q-networks), hierarchical decomposition, and distributed computation can reduce memory complexity from $O(|\mathcal{S}||\mathcal{A}|^2)$ to $O(|\Theta|)$, where $|\Theta|$ is the number of neural network parameters. These approaches enable real-time decision support while maintaining convergence guarantees.

\section{Conclusion}

The experimental validation demonstrates the effectiveness of game-theoretic approaches for cyber-physical microgrid resilience, offering insights for both theoretical understanding and practical deployment.

\subsection{Key Observations}

\textbf{Adaptive Strategy Superiority:} Adaptive defense strategies outperform static methods, with gains from 12.4\% (Q-Learning) to 35.5\% (Stackelberg-RL). This reflects the mismatch between static Nash assumptions and dynamic adversarial environments, supported by the $\mathcal{O}(1/\sqrt{T})$ convergence rate.

\textbf{Information Structure Impact:} Stackelberg-RL achieves the highest score (0.9919) by modeling sequential decision-making where defenders act first. The minimax formulation $D^* = \argmax_j \min_i M_{ij}$ guarantees robust security, as seen when $D_4$ fully neutralized $A_1$.

\textbf{Bounded Rationality Benefits:} Softmax and regret matching improve resilience (20.6\% and 28.1\%) by incorporating probabilistic and history-based adaptations. Softmax temperature ($\tau=0.5$) captures bounded rationality, while regret bounds ensure reliable convergence.

\textbf{Multi-Agent Learning Challenges:} Multi-agent Q-learning offers modest gains (14.6\%) but faces instability from non-stationarity and heterogeneous attack–defense interactions, highlighting practical implementation challenges.

\subsection{Practical Implications}

Payoff matrix analysis shows that tie-line switching ($D_2$, $D_4$) and combined strategies ($D_{10}$) provide strong resilience across diverse threats, suggesting priority areas for microgrid investments. Scalability studies confirm feasibility: 33-bus analysis completes in 2.3 hours, while 69- and 123-bus remain practical. Monte Carlo validation further provides confidence bounds for operational use.

\subsection{Limitations and Future Directions}

Despite strong results, limitations remain. Current attack models exclude advanced persistent threats or AI-driven adversaries, and discrete strategy spaces may miss continuous dynamics. State space growth challenges scalability, motivating function approximation and distributed solutions. Validation is limited to a single testbed; broader network types and attack vectors should be explored. Deployment challenges also exist in SCADA/EMS integration, regulatory compliance, and operator training.

Future work will extend to advanced threat modeling, multi-microgrid coordination, uncertainty-aware optimization, and integration with real-time EMS/SCADA systems. These directions will enhance scalability, realism, and acceptance for field deployment.

Overall, the results confirm that game-theoretic and adaptive learning frameworks significantly enhance cyber-physical microgrid resilience while identifying clear opportunities for refinement and extension.

\bibliographystyle{IEEEtran}
\bibliography{references}
\end{document}